# Acoustic Zero-Index Metamaterials for Leaky-Wave Antennas

Keqiang Lyu[1], Ying Wu[1,2]*

[1]Division of Physical Science and Engineering, King Abdullah University of Science and Technology (KAUST), Thuwal 23955-6900, Saudi Arabia

[2]Division of Computer, Electrical and Mathematical Science and Engineering, King Abdullah University of Science and Technology (KAUST), Thuwal 23955-6900, Saudi Arabia

*Email: Ying.Wu@kaust.edu.sa

In this work, we introduce an advanced acoustic leaky-wave antenna employing zero-refractive-index metamaterials (ZIMs) that significantly surpasses traditional designs in terms of radiation efficiency and directivity. Our design features a novel space-coiling structure, which manipulates the Dirac-like dispersion based on accidental degeneracy to achieve double-zero-index metamaterials (DZIM). The newly developed acoustic Dirac leaky-wave antenna (ADLWA) achieves more than double the radiation efficiency of conventional antennas based on arrays with side holes and membranes. It exhibits superior directional control, allowing for precise beam scanning by adjusting the frequency. Additionally, the ADLWA functions effectively as a passive sonar system, capable of detecting the direction of moving sound sources. This breakthrough enhances acoustic wave control and promises significant advancements in sensing and communication technologies.

## I. Introduction

Acoustic leaky wave antennas (ALWAs) are passive, dispersive waveguides that have been extensively explored for their capability to achieve frequency-dependent directional radiation scanning. Generally, conventional ALWAs are composed of waveguiding materials with periodic side holes, supporting only right-hand (RH) propagation and allowing forward radiation [1]. Composite Right/Left-Hand Transmission Line (CRLH-TL) metamaterials [2,3], which enable both right-hand (RH) and left-hand (LH) propagation to facilitate the transition of the main radiation beam from backward to forward radiation, have been introduced into ALWAs [4]. In prior research, by incorporating periodic arrays of side holes and membranes along the waveguide [5,6], the design of these CRLH-TL ALWAs enables the wavenumber to shift continuously from negative to positive, without the presence of stopband. Many studies based on this structure were expanded [1,7–15], such as acoustic dispersive prisms [12] and single-microphone acoustic source direction finding [13,14]. More recently, heterostructures [14] and gradient-distributed structural units [15] have been designed to achieve higher directivity. Additionally, underwater acoustic leaky-wave technologies have been introduced [16–18]. However, LWA structures with side holes and membrane arrays often face issues such as complex assembly, challenging membrane tension control, high losses, and large side lobes. Their maximum radiation efficiency ranges from a few percent [12] to 25% [4].

Zero-refractive-index metamaterials (ZIMs) are a class of unconventional materials characterized by a refractive index of zero [19–24]. In ZIMs, the phase velocity becomes infinite, resulting in no spatial phase change. This exhibits remarkable properties such as super-coupling [25–27], cloaking effects [28–31], and wave front engineering [32–34]. Recently, ZIMs have demonstrated a unique beam-collimating effect, producing highly directional beams in the far field [35–37]. This material overcomes the limitations of conventional antennas, where directivity is typically constrained by the nonuniform spatial phase distribution across the radiation aperture, ultimately approaching the fundamental directivity limit of previous antenna designs [35,37]. This advancement opens new avenues for enhancing acoustic antenna technology. Therefore, integrating ZIMs with ALWAs holds significant potential for achieving higher radiation efficiency and improved di-rectivity. To achieve ZIMs, various approaches have been explored, including the use of membranes [26,38], Helmholtz resonators [39–41], and waveguides [25,42]. In optics, lossless double-zero-index metamaterials (DZIMs) are typically realized using photonic crystals composed of high refractive index scatterers that exhibit a Dirac-like cone induced by accidental degeneracy, resulting in both zero permittivity and permeability [20]. However, realizing a similar mechanism directly in air remains challenging due to the significantly slower speed of sound in air compared to solids and liquids. Prior acoustic DZIMs have primarily been developed in water [27,43], elastic wave [23], high-order waveguides [44], flat dispersion [30], Helmholtz resonators [40], or 3D configurations [45,46]. In recent years, the space-coiling structure demonstrates a high capacity for novel acoustic wavefront manipulation by effectively increasing the refractive index through air folding and prolonging sound propagation time [47–51]. We speculate that this structure could be an ideal candidate for realizing acoustic DZIMs in 2D air media.

In this letter, we introduce a novel space-coiling structure with variable channel spacing that effectively disrupts the inherent multipole degeneracy of traditional designs [52]. Through precise manipulation, we achieved a low-frequency subwavelength Dirac-like point at the center of the Brillouin zone, accompanied by the manifestation of double zero effective parameters. The DZIM is arranged periodically in a one-dimensional waveguide, forming an acoustic Dirac leaky-wave antenna (ADLWA). This newly designed ADLWA demonstrates the ability to radiate through broadside while dynamically scanning the directive beam by frequency variation. Distinctively, the ADLWA features a closed stopband, matched response, and a consistent leakage constant around broadside, significantly enhancing its directivity and more than doubling its radiation efficiency relative to conventional LWAs. Moreover, this ADLWA also function as a passive sonar system, adept at locating moving sound sources in space independently of its physical length.

## II. Design of the DZIMs

Acoustic multiple Mie resonance has been found to occur in space-coiling structure [47], which provided novel abilities of wave manipulation by generating a negative bulk modulus with a negative mass density at resonance. It was found in [52] that in this intriguing type of resonance, Mie resonance frequencies of various orders overlap, that is, degenerate Mie resonance. This mechanism is enabled by the anisotropic character of the effective mass density. It has been shown that the accidentally degenerate Dirac cone of dipole mode and monopole mode at the Γ point can be mapped to a double zero [20].

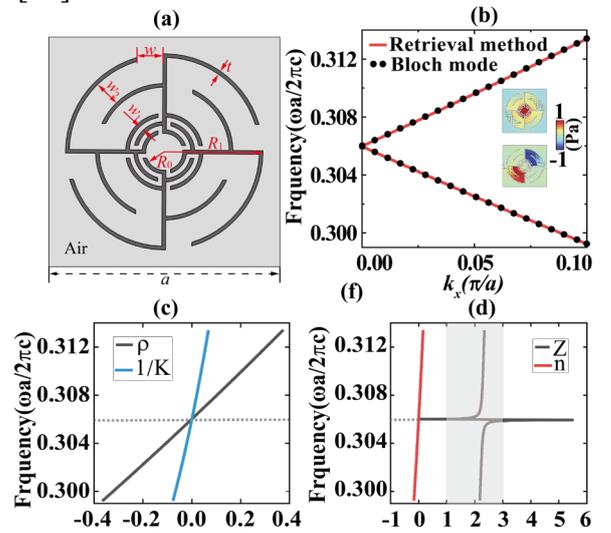

FIG. 1. Schematic diagram of the designed space coiling structure and the corresponding properties of the metamaterial. (a) The structure with four zigzag channels. The rigid wall (thin solid lines) inserted into air (gray color) forms a zigzag channel with opening width and outer channel width $w_2 = 0.114a$, inner channel width $w_1 = 0.02a$, inner radius $R_0 = 0.08a$, and outside radius $R_1 = 0.427a$. (b) Comparison of band structure of DZIM computed by determining the angular frequencies as a function of wave vector for all the Bloch modes (black dot), and by $k = n_{eff}\omega/c$ with the retrieved effective index $n_{eff}$(red curve). The inset is eigenmodes near the Dirac-like point with a small $k$ along ΓX direction. The effective (c) mass density $\rho_{eff}$ (black line), reciprocal of bulk modulus $1/\kappa_{eff}$ (blue line), (d) impedance $Z$ (grey line) and refractive index $n_{eff}$(red line) as a function of frequency near the Dirac-like point.

In this work, we enhance traditional space-coiling metamaterials by introducing variable channel spacing, unlike the conventional constant width. This innovative design significantly enhances the tunability of the material's equivalent constitutive properties, thereby offering a more flexible approach to the customization of its acoustic responses. As shown in Fig. 1(a), the lattice constant is denoted by $a$, hard solid plates with thickness $t$ are embedded into the background fluid, creating zigzag channels. The width of each channel is treated as a design variable, allowing for precise manipulation of the channel geometry. In this work, we engineer the unit cell to induce accidental degeneracy between the dipolar and monopolar modes. Considering that the acoustic Mie resonance monopole fields are primarily localized near the center, while dipole fields are localized in regions farther from the center, as illustrated in Fig. 1(a), the varying widths of the inner and outer channels can independently adjust the positions of monopole and dipole frequencies until degeneracy, or even effectively eliminating the influence of quadrupoles.

Fig. 1(b) shows that along the $\Gamma X$ direction, two branches with linear dispersion intersect at the degeneracy point $k = 0$, forming a Dirac-like point at $\omega = 0.306(2\pi v/a)$. Notably, this represents a triply degenerate state, including a monopolar mode as depicted in the inset, a dipolar longitudinal mode along the $\Gamma X$ direction, and a transverse mode forming a flat band (as shown in Fig. S1 in supplementary). This degeneracy arises accidentally, thus, altering the geometric parameters can split this triply degenerate state into a doubly degenerate state and a single state. The cone dispersion can be mapped to DZIM, where the reciprocal of the effective bulk modulus $1/\kappa_{\text{eff}}$ and the effective mass density $\rho_{\text{eff}}$ both approach zero. Fig. 1(e) illustrates the effective mass density $\rho_{\text{eff}}$ (black solid line) and the reciprocal of the bulk modulus $1/\kappa_{\text{eff}}$ (blue solid line) as a function of frequency, based on field averaging [53,54], intersecting at zero at the Dirac point frequency $\omega = 0.306(2\pi v/a)$. At this frequency, the structure also exhibits an effective refractive index near zero and an effective impedance close to that of the background air, as shown in Fig. 1(d). To quantitatively evaluate whether the effective constitutive parameters are valid, we obtain the dispersion curve as ($k = n_{eff}\omega/c$) (red solid line in Fig. 1(b)), based on $\rho_{\text{eff}}$ and $1/\kappa_{\text{eff}}$. The close correlation between the analytical and numerical results confirms that the effective homogeneous medium approximation accurately describes the metamaterials composed of space-coiling structure near the Dirac point frequency for this configuration.

### III. Acoustic Dirac Leaky wave antenna (ADLWA)

To facilitate effective coupling between the metamaterial and propagating waves in air, an interface between the metamaterial and air must be designed to support the propagation of leaky modes. Crucially, this propagation needs to persist at the $\Gamma$-point without a frequency gap [55]. This enables continuous scanning from backward to forward radiation, leading to the formation of the ADLWA. According to previous work [18-20], short shunt tubes can be used between background air and waveguide in traditional acoustic LWAs to create a leaky interface, depending on the specific geometry requirements.

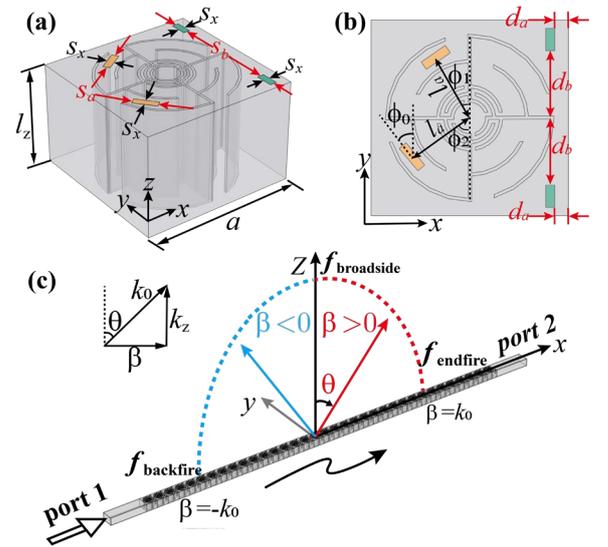

FIG. 2. The leaky Dirac metamaterial. (a) Unit cell of one-dimensional Dirac metamaterials with top perforations featuring 4 slots (yellow and green rectangles) in the acoustic waveguide. The periodicity is in the x-direction. (b) The top view of the unit cell shows the location

of the slots. (c) A three-dimensional view of the 35-cells ADLWA, formed by cascading unit cells of the leaky Dirac metamaterials.

Here, we employ a more straightforward leakage mechanism by considering the top-perforated unit cell, as shown in Figs. 2(a) and (b). Notably, the introduction of perforations disrupts and shifts the original Dirac point, as compared to the unperforated DZIM in the waveguide. Numerical results show that by fine-tuning the geometry of the radiation apertures, the two orthogonal Bloch eigenmodes of the original DZIM within the waveguide can re-degenerate. These apertures, which can vary in shape, being either circular or rectangular, are strategically placed to maintain the waveguide's ability to support its original modes. Specifically, the size of the slot can adjust the imaginary part of the eigenfrequency, that is, the leakage factor of the leakage mode unit. As shown in Fig. 2(b), the left slot (yellow area) and right slot (green area) are used to adjust the eigenfrequency of the dipole mode and monopole mode separately. The relevant geometric parameters are $\phi_0 = 37°$, $\phi_1 = 30°$, $\phi_2 = 55°$, $d_a = 0.07a$, $d_b = 0.343a$, $s_x = 0.0456a$, $s_a = 0.148a$, $s_b = 0.114a$, $l_z = 0.656a$. The 2 leaky modes have finite (due to radiation) and comparable quality factors $Q_d = 150.6$ and $Q_m = 94.9$. The ADLWA presented in Fig. 2(c) features a 3D view of the 35 cells. This antenna is constructed by directly cascading the unit cells of the previously described leaky Dirac metamaterials. As the frequency of the incident wave varies, the group velocity and the propagation constant of the metamaterial shifts from negative to positive. This enables it can continuous radiation scanning from backward to forward in the $xz$-plane. In other words, the phase constant $\beta$ along the $x$-direction determines the primary radiation angle $\theta$ for the DLWA. This relationship can be expressed as

$$\theta(f) = sin\frac{\beta(f)}{k_0} \quad (1)$$

Where $k_0$ is the wave vector of the ambient medium.

Leaky wave antennas are commonly analyzed using transmission line theory (TLs), which is a circuit-based concept that can describe wave propagation across various systems [14,56]. We utilize a circuit model to analyze our leaky metamaterials. It should be noted that the typical one-to-one correspondence [12–21] between structure and circuit components may not be readily apparent in this structure. As illustrated in Fig. 3a, distinct series and shunt resonators were used, akin to the modeling approach in [15], to represent different resonant modes. Each mode is effectively modeled by either an inductor/capacitor resonator, ensuring that the circuit elements corresponding to these resonances remain distinct. This unit cell circuit model decouples the two resonances, mirroring the orthogonal eigenmodes of the metamaterial unit. Since each mode exhibits a finite Q-factor due to radiation losses, the circuit model includes both lossy series and shunt resonators, represented by resistor/LC (RLC) or conductance/LC (GLC) configurations. The corresponding series impedance $Z_{se}$ and parallel admittance $Y_{sh}$ are characterized as follows:

$$Z_{se} = R_{se} + j\omega L_{se} + \frac{1}{j\omega C_{se}} \quad (2)$$

$$Y_{sh} = G_{sh} + j\omega C_{sh} + \frac{1}{j\omega L_{sh}} \quad (3)$$

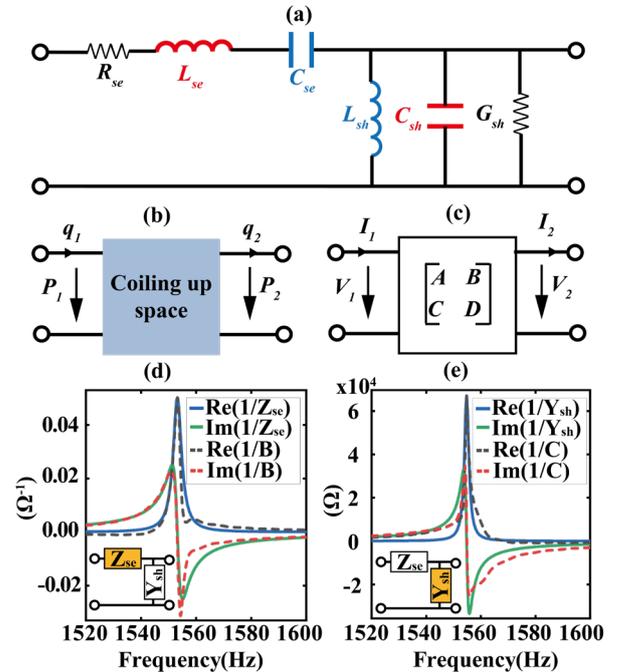

FIG. 3. (a) Illustrate the schematic diagram and the equivalent acoustic transmission line circuit for a single unit. (b) Two-port acoustic transmission of a metamaterial unit. (c) Characterization of a two-port network using the [ABCD] matrix in a circuit mode. In the 35-unit

simulation, characterization of series and parallel loss resonance in unit cells includes: (d) extracting the reciprocal of the B element from the unit transfer matrix ABCD near the $\Gamma$ point (dashed line) and fitting the curve of the circuit model $Z_{se}^{-1}$(solid line); (e) extracting the reciprocal of the C element from the unit transmission matrix ABCD near the $\Gamma$ point (dashed line) and fitting the curve for the circuit $Y_{sh}^{-1}$ (solid line).

This straightforward model offers several key insights into the behavior of ADLWA, such as verifying that the stopband is closed and that a real, constant Bloch impedance near the broadside frequency enables matched operation. During resonance, both the impedance and admittance become purely real, specifically, $Z_{se} = R_{se}$, $Y_{sh} = G_{sh}$. Consequently, the resonance frequencies, $f_{se}$ and $f_{sh}$, are defined as follows:

$$f_{se} = \frac{1}{2\pi\sqrt{L_{se}C_{se}}} \quad (4)$$

$$f_{sh} = \frac{1}{2\pi\sqrt{L_{sh}C_{sh}}} \quad (5)$$

In our acoustic circuit modeling, voltage corresponds to acoustic pressure $p$ and current to volume velocity $q$. We suggest that the transmission properties of a single unit cell from the leaky Dirac metamaterials, as shown in Fig. 3(b) (based on full-wave simulations), can be effectively modeled by the ABCD matrix of a two-resonator circuit at the degeneracy frequency, as depicted in Fig. 3(c). This approach enables the use of a simple circuit model to represent the two-port behavior of the unit cell, thereby elucidating its dispersion and matching characteristics. The two-port S-parameters of the 35-cell antenna can be obtained through full-wave simulations in COMSOL, and subsequently converted to an $ABCD$ transmission matrix. To derive the ABCD matrix of a single cell, we use $ABCD_1 = \sqrt[35]{ABCD_{35}}$, where $ABCD_1$ and $ABCD_{35}$ represent the matrices for a single unit cell and the 35-cell structure, respectively. We then determine critical parameters such as phase ($\beta$) and leakage ($\alpha$) constants, Bloch impedance ($Z_b$) and other relevant circuit model parameters from $ABCD_1$. For structures with 45 and 55 cells, the matrices $ABCD_1 = \sqrt[45]{ABCD_{45}}$ and $ABCD_1 = \sqrt[55]{ABCD_{55}}$ are calculated similarly. In transmission line theory, when there are periodic circuit components, the ABCD matrix of a circuit having both a series impedance ($Zse$) and a shunt admittance ($Ysh$) is [4,15,57]:

$$ABCD_{se\times sh} = ABCD_{se} \times ABCD_{sh} = \begin{bmatrix} 1+Z_{se} & Z_{se} \\ Y_{sh} & 1 \end{bmatrix} \quad (6)$$

This indicates that analyzing the B and C elements of the ABCD matrix for this circuit separately reveals the series impedance and shunt admittance, respectively. Subsequent examination of the $ABCD_1$ matrix is detailed in Fig. 3(d) and 3(e), which plot the complex values of the inverse B and C elements across frequencies (represented as red and blue dashed lines, respectively) for the purpose of model extraction. The presence of distinct resonances corresponding to each matrix element solidifies the validity of our series/shunt model, effectively encapsulating the two-port functionality of the unit cell in a coherent manner. As noted, frequencies at which B and C's imaginary parts reach zero define the resonant frequencies ($f_{se}$ and $f_{sh}$), and their real components are set as $R_{se}$ and $G_{sh}$. Using curve fitting, we calculate the inductances ($L_{sh}$ and $L_{se}$) and model $C_{sh}$ and $C_{se}$ as functions of these inductances, according to Eqs. (4) and (5). As illustrated in Figs. 3(d) and (e), the resonance parameters modeled using transmission line theory and extracted via full-wave simulation show a strong correlation. The extracted resonances ($f_{se} \sim f_{sh} \sim$ 1554 Hz) and the fitted model parameters ($R_{se}$ =19.0912 $\Omega$, $L_{se}$ =0.85 H, $C_{se} = 1.234 \cdot 10^{-8}$ F, $G_{sh} = 1.491 \cdot 10^{-5}$ $\Omega^{-1}$, $L_{sh} = 8.8 \cdot 10^{-3}$ H, $C_{sh} = 1.192 \cdot 10^{-6}$ F) verifies that the structure's stopband is conclusively closed, indicating a balanced condition. The propagation of acoustic waves through the periodic structure is analyzed using Bloch theory, which is based on the unit cell. The Bloch propagation constant $\gamma$ is defined as follows [1,57]:

$$\gamma = \frac{1}{a}arcosh\left(1 - \frac{Z_{se}Y_{sh}}{2}\right) = \alpha + j\beta \quad (7)$$

Here, the $Z_{se}$ and $Y_{sh}$ in TL model correspond to the B and C elements of the metamate-rial unit's acoustic field transmission matrix, respectively. As depicted in Figs. 4(a) and (b), the leakage factor $\alpha$ and phase constant $\beta$ near the $\Gamma$-point, as functions of frequency for the fabricated ADLWA, were determined using S-parameter retrieval from the driven N-cell configuration.

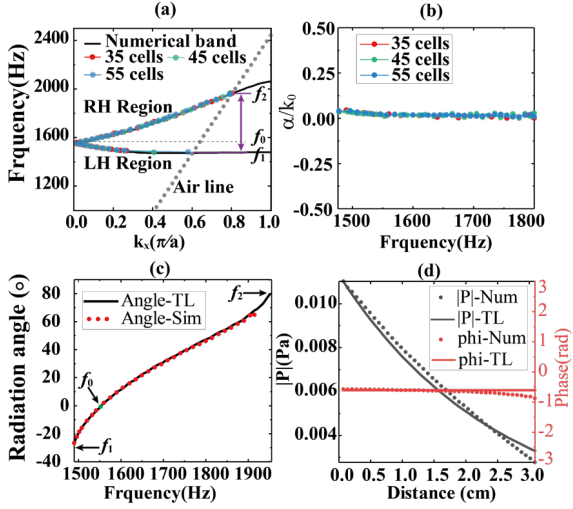

FIG. 4. The propagation constants of the DLWA are determined, with (a) the dispersion diagram extracted from the N-cell simulation illustrating a closed broadside stopband, and (b) the normalized attenuation (leakage) constant $(\alpha/k_0)$ around broadside frequencies, which exhibits minimal variation. (c) The radiation beam angle varies as a function of the incident frequency. Solid curves represent the TL beam angles for 35 cells, while dotted lines indicate the simulated beam angles for the DLWA. (d) The aperture field and phase distribution along the length of the antenna are sampled at the center of the slots for broadside radiation.

The radiation region of the ADLWA is located above the sound line, represented by the black dashed line, indicating radiation in the fast-wave regime. The $\beta = 0$ (zero-index) occurring at $\Gamma$ point, which is 1554 Hz ($f_0$), along with the non-zero group velocity, confirms the viability of broadside radiation. Fig. 4(b) reveals that the leakage constant $\alpha/k_0$ is approximately 0.0267 at $f_0$ and remains stable, deviating from the usual patterns observed in LWA [1,11,12,15]. Typically, LWAs exhibit pronounced fluctuations in the leakage constant around broadside due to phenomena such as open stopbands mechanisms [58]. When excita-tion occurs at frequencies within the left-handed (LH) region in TLs, between $f_1$ and $f_0$, negative group velocity results in backfire radiation. Conversely, forward radiation happens in the right-handed (RH) region, between $f_0$ and $f_2$. The Bloch parameters $\beta$ are employed to determine the radiation angle of the LWA. The main beam angles $\theta$ at various frequencies are calculated using the finite element method (FEM) or Eq. (1) from TL theory. Fig. 4(c) displays a comparison of the results obtained from these two methods for 35 cells. The frequency range presented is confined to the radiating band, spanning from $f_1 = 1490\ Hz$ to $f_2 = 1955\ Hz$. The above shows the ADLWA lacks a stopband and seamlessly transitions from the backward to the forward region without any gap. Furthermore, the phase and magnitude functions of the radiation pattern for the N-celled array-type LWA are denoted as $\varphi_n$ and $P_n$, respectively:

$$\varphi_n = -(n-1)k_0 a * \sin\theta \quad (8)$$

$$P_n = P_0 e^{-\alpha(n-1)a} \quad (9)$$

Where $P_0$ is the input pressure at the first unit cell element. The zero-index material (ZIM) behavior is characterized by a zero-phase progression along the length of the antenna, which results in broadside radiation when the antenna is excited unilaterally with minimal reflected waves. In Fig. 4(d), the red solid line represents the phase distribution as calculated by Eq. (8). Meanwhile, the red dotted line is derived from simulation calculations of the field's phase, sampled at the center of the slots, for the frequency corresponding to the $\Gamma$-point of the DLWA with 35 cells. The agreement between these two lines confirms the zero-phase propagation phenomenon. Additionally, the exponential decay of the field magnitude, depicted by the black solid line (TL model) and the black dashed line (simulation), further confirms the unilateral leaky operation characteristic of zero-index propagation.

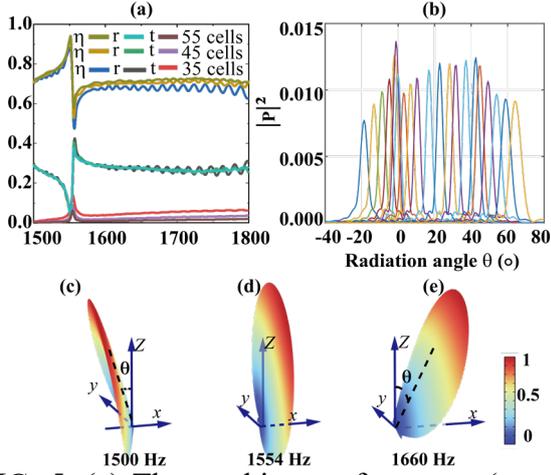

FIG. 5. (a) The working performance (transmission, reflection and radiation coefficient) of ADLWA with different lengths. (b) Scanning capability of the ADLWA accompanied by significantly low side lobes. The far-field radiation pattern of the ADLWA reveals backward radiation (c) at 1500 Hz, broadside radiation (d) at 1554 Hz, and forward radiation (e) at 1660 Hz

Figure 5. (a) shows the working performance of ADLWA with different lengths. The radiation efficiency $\eta$ of the ADLWA showing a remarkable performance. What's also interesting is that our radiation efficiency is around 70% throughout the entire back-to-forward scanning process. The efficiency of all previous acoustic leaky wave scanning antennas composed of membranes and acoustic cavities, whether uniformly arranged or specially designed with gradients, is less than 25%. Here, the radiation efficiency of our structure is more than twice that of previous acoustic leaky wave scanning antennas. Fig. 5(b) demonstrates the DLWA's extensive scanning or steering capabilities. The sidelobe levels consistently stay well below the radiation power peak across all frequencies, reflecting its precise control over wave directionality. Fig. 5(c-e) presents the far-field radiation patterns (normalized power distribution) of the structure at three distinct frequencies: 1500 Hz, 1554 Hz, and 1600 Hz, corresponding to radiation in the backward, broadside, and forward directions, respectively. Additionally, we found that the DZIM offers us diverse modulation capabilities. Traditional acoustic waveguide leaky wave antennas are constrained by complex impedance matching issues, making them inflexible for adapting to specific requirements in complex scenarios. Leveraging the super-coupling effect of DZIM, our ADLWA can be customized to achieve arbitrary radiation patterns by altering the antenna shape (as shown in the supplementary material). Furthermore, even when partial antenna elements are damaged, the overall super-coupling effect ensures stable operation of the antenna (as shown in Fig. S8 in the supplementary material).

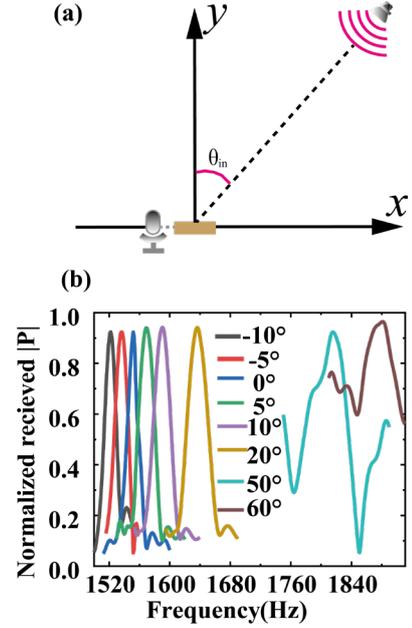

FIG. 6. (a) The schematic diagram of the ADLWA for localizing the acoustic source. (b) Received sound pressure spectra captured by the microphone in the DLWA for various incident angles $\theta$.

We know that single sensor positioning has become a research focus in recent years. This intriguing approach could potentially disrupt traditional methods that depend on multi-sensor beamforming to pinpoint sound source locations. We have studied the ability of leaky wave antennas to steering sound waves in a frequency-dependent direction. According to the reciprocity principle, when operated in reverse, the antenna exhibits a well-defined frequency-direction relationship. In other words, ADLWA can be regarded as a passive sonar, which can detect the moving sound source in the space from the radiating side. As illustrated in Fig. 6(a), a microphone is positioned at one end of the antenna within a designated yellow region.

A point sound source Q, emitting at frequency $f$, is located in the far field. The angles between the sound source and the antenna are denoted by $\theta$. It is critical to ensure that the distance from Q to the antenna is significantly greater than the shortest wavelength in the operational frequency band, allowing us to disregard any phase differences by the time the sound waves reach the microphone. The antenna captures the sound pressure spectrum as the angles between it and the sound source vary. During this process, the received sound pressure spectrum is expected to exhibit a peak at a specific direction. As depicted in Fig. 6(b), the sound pressure spectra recorded at the microphone position for each incident angle reveal a lobe with maximal amplitude at a specific frequency. Using Eq. (1), the angle $\theta$ can be determined, thereby confirming the direction of the sound source Q.

## IV. CONCLUSION

In summary, we designed a novel double-zero refractive index metamaterial (DZIM) in air, utilizing a space-coiling structure, and successfully verified its zero-refractive-index properties. By integrating this metamaterial with a waveguide, we developed a highly efficient acoustic Dirac leaky-wave antenna (ADLWA) based on DZIMs, which enables continuous beam scanning from backward to forward radiation without a stopband. The ADLWA exhibits superior radiation efficiency and enhanced directivity compared to conventional leaky-wave antennas. The designed ADLWA serves a dual purpose, functioning as both a radiating antenna and a passive sonar system, offering promising prospects for the development of multifunctional, high-performance acoustic devices. This work lays a solid foundation for the further exploration of Dirac-like acoustic metamaterials in practical and adaptable applications.

## V. ACKNOWLEDGEMENT

The research reported in this manuscript was supported by King Abdullah University of Science and Technology Baseline Research Fund.